\documentstyle[epsfig]{elsart}
\begin{document}
\begin{frontmatter}
\title{Measurements of the radiation hardness of selected
  scintillating and light guide fiber materials}
%\thanks

\author{E.C. Aschenauer},
\author{J. B\"ahr}
{\renewcommand{\thefootnote}{\fnsymbol{footnote}}
\hspace{-0.3cm}\footnote{\small corresponding author, phone: +49 33762
    77235, Fax: +49 33762 77330,\\ \hspace*{0.3cm} e-mail:
    baehr@ifh.de}},
\author{R. Nahnhauer},
\author{R. Shanidze\thanksref{tbilisi}}
\address{DESY Zeuthen, 15738 Zeuthen, Germany}
{\setcounter{footnote}{0}
\thanks[tbilisi]{on leave from High Energy Physics Institute, Tbilisi
  State
University, Georgia}}

\author{D. Fink, K.H. Maier, M. M\"uller}
\address{Hahn-Meitner-Institut, Glienicker Str. 100, D-14109 Berlin, Germany}

\author{H.A. Klose} 
\address{Ingenieurb\"uro OPTOMET,D.-Brinkmann-Str. 39B, D-26125 Oldenburg, Germany}

\author{M. Sprenger}
\address{Lambda Physics, Hans-B\"ockler-Str. 12, D-37079 G\"ottingen, Germany}

%, P. Goppelt-Langer}
%\address{Gesellschaft f\"ur Mess- und Systemtechnik, Rudower Chaussee
%  5, D-12489 Berlin, Germany}

%\thanks

\begin{abstract}

Radiation hardness studies of KURARAY SCSF-78M scintillating fibers
and clear fibers from KURARAY and pol.hi.tech.
% are summarized. They were 
 performed under
different dose rate conditions in proton and electron beams are
summarized. For high
dose rates in-situ measurements of the fiber light output were 
done. During several months after irradiation all fibers were measured
concerning light emission and transparency. 
 
Fibers irradiated at high rates to about 1 Mrad are clearly damaged but recover
within a few hours up to several weeks. Using smaller  rates up to
the same integral dose a decrease of the light output of scintillating
fibers of up to 30\,$\%$ can not be excluded. Clear fibers seem to be
uneffected up to 400 krad.
No significant influence of fiber coverage and atmosphere during
irradiation was found.  

\end{abstract}
\end{frontmatter}

\section{Introduction}

Recently \cite{lit1}-\cite{lit3} a fiber detector was developed as
an alternative solution for the inner tracker of the HERA-B experiment
\cite{lit4}. With an 
accelerator cycle of 96\,nsec and  four events per cycle with a charged
multiplicity of about 200\,the detector modules have to work
 several years under
% non-negligible dose rate conditions. The
an estimated
integral dose per year in the inner tracker region of about 1\,Mrad.

 The light produced by particles crossing the scintillating fibers of
 the detector is transported by 3\,m long light guide fibers to 64
 channel multianode photomultipliers Hamamatsu R5900-M64\footnote{
 Hamamatsu Photonics K.K., Electron tube division, 3124-5, Shimokanzo,
 Tokyooka Village, Iwatagun, Shizuoka-ken, Japan}
 available only
 with bialcali  photocathodes.
% A preselection of materials gave
 Best
% results for
 light output and long term stability were obtained for KURARAY\footnote
{KURARAY Co.Ltd., Nikonbashi, Chuo-ku, Tokyo 103, Japan}
 scintillating double clad fibers SCSF-78M and clear double clad fibers
 from KURARAY and pol.hi.tech.\footnote{pol.hi.tech., s.r.l., S.P.
   Turanense, 67061 Carsoli(AQ), Italy}. The
 corresponding radiation
 hardness studies were performed with high dose rates in 70\,MeV proton
 and 2\,MeV electron beams of the Hahn-Meitner Institute 
Berlin
%\footnote{Hahn-Meitner Institut Berlin, Glienicker Str. 100,
%  D14109 Berlin}
 \cite{lit1},\cite{lit3},\cite{lit5},\cite{lit6}
 and photons from a $^{60}$Co source \cite{lit2}. In the first case in-situ
 measurements of the light output even with spectral resolution were
 possible.

In the past there were several arguments \cite{lit7}-\cite{lit9} that the presence of
oxygen during and after the irradiation may be important for the
observed damage. In this case also the dose rate may influence the
final result because diffusion processes are time dependent. 

Our results from high dose rates using charged particle beams will be
summarized below and compared to new data from low dose rate exposures
of the same materials. The new tests were performed in air and
nitrogen atmosphere with glued and non-glued scintillating fibers
and compared with non-irradiated test samples.
% have been measured every time for comparison. 

\section{Experimental conditions}

\subsection{Fiber samples}
The fiber samples for proton and electron irradiation have 
the same global structure as shown in Fig. 1.
 For high rate  irradiation 4$\times$4 fibers of 0.48\,mm diameter were
 glued together
resulting in a cross section of about 2$\times$2\,mm$^2$. The samples for
electron low dose rate irradiation consist of a fiber arrangement of
1$\times$7 fibers of the same diameter forming a fiber road in the
later detector.\
Coupling pieces are mounted at both ends of the
30\,cm  long samples  in which the ends of the fibers
 are inserted, glued and polished.
This allows an optical coupling to light guides or photomultipliers with
 light losses of less than 10\,\%. The fiber samples are mechanically stabilized by two
brass rods of 3\,mm diameter.\

 For low dose rate electron irradiation there were two types of samples. The
first type is fully glued to shield the fibers from the gaseous
environment, whereas the second type is mounted using a minimum of
 glue in thin strips near the connectors in
order to allow the gaseous atmosphere to have contact to the fiber material.

For the in-situ measurements single fibers were coupled at one or both
ends to glass fibers which transport the light to the corresponding
spectrometers (see Fig. 2).

\subsection{Irradiation setup}

A schematic view of the irradiation setup in the proton and
electron beams is given in Fig. 3. For electron irradiation the beam was
extracted from the vacuum 
system through a window of 100\,$\mu$m thick Aluminium and 40\,$\mu$m Inconel.
 A metallic aperture of
3$\times$12\,mm$^2$ was used for beam profile definition.
 In the case of the proton irradiation
the beam was extracted
through a 7\,$\mu$m thick Tantalum foil. The beam size and the emittance
angle were limited by two PMMA  (polymethyle methacrylate)
apertures. The total range of protons
in fiber material is about 39\,mm which is checked by the profile of
the colour changes in the PMMA  aperture 
during the irradiation. The
spot size and position was additionally monitored by polyvinylalcohol
(PVA) methylene blue plastic detector foils \cite{lit10}. The dye is
radiation sensitive and its degradation yield is proportional to the
irradiated particle fluence. The degradation of the dye in the foil has
been determined by UV-VIS-spectroscopy.  A typical result of such beam
homogeneity control for proton irradiation is depicted in Fig.
4. The higher the transparency the higher was the irradiation dose in
the given area. At positions 1, 2 a high radiation level with low
restriction in the field distribution is registrated. The profile
created by the plastic apertures is given by curve 3 and 4. A low
non-structured irradiation level is characterized by curves 5 and 6.

For the in-situ registration of beam excited scintillation spectra
fiber optic PC-plug-in spectrometers (Ocean Optics\footnote{Ocean
  Optics Inc., 380 Main Street, Dunedin, FL34698, USA})
 were used. They were placed outdoors of the cave in order to    
%of 22\,m long light-guiding glass fiber cables outdoors of the cave in
 suppress the high radiation background using 22\,m long light-guiding
 glass fibers . A detailed
description of the used experimental setups in both cases can be
found in \cite{lit5}.

The proton irradiation was performed quasi point-like at two points
 along the sample with
a dose of $\ge$1\,Mrad at 20\,cm and of 0.1\,Mrad at 10\,cm
respectively within a few minutes. The dose rate was
about 30\,Mrad/h. The irradiation of short areas of the samples gave
the possibility  to
separate the damage of scintillator and optical matrix.

The same irradiation procedure was applied for in-situ measurements 
of radiation
damage using a corresponding electron beam.

High current irradiations were only carried out under ambient
atmosphere using cooling by a powerful fan. The temperature rise
during the irradiation could be neglected \cite{lit5}.

A new series of tests has been performed irradiating the fiber
material with a relatively low dose rate of 2 MeV electrons to
approximate the later experimental conditions. A  dose of about 1\,Mrad
was applied during five periods within about nine weeks.
The particle flux was monitored by a
matrix of Faraday cups. The distance between scatter foil and sample
plane was about 1.5\,m.
In this case the samples were kept
either in air or in nitrogen atmosphere.
% Non-irradiated
%reference samples were used for comparison with irradiated samples to
%minimize systematic errors.
\\

\subsection{Measurement procedure}

In-situ registration of scintillation spectra first described in \cite{lit11}
was performed for the beam
excited regions 1 and 2 (see fig. 2). The spectra were measured
during the whole irradiation time in the first irradiated region 1 of
fibers and after that in the second region 2 under influence of high
absorption in the presumably predamaged region 1 in order to determine
  the change
of the absorption coefficient during the irradiation. Between
irradiation procedure 1 and 2 a preparation time of a few minutes was
necessary. The beam excited scintillation spectrum served in the
second case as changeable light source for absorption measurements in
a limited spectral region.

For recovery measurements in the laboratory a few hours after
irradiation the optical excitation was realized by a high pressure
Hg-lamp at $\lambda$ = 365\,nm
% using a monochromator or the fiber-optic spectrometer
. \\
In addition to the in-situ measurements which used single fiber
samples and UV-excitation for the measurement in the laboratory 
%after the irradiation
investigations were done using multi-fiber bundles. 
%were used for the irradiation and
%measurements which were not realized in-situ.
The irradiated multi-fiber samples (see section 2.1) were evaluated using a $^{106} Ru$ source. The fiber sample was mounted within a source
collimator slit. The light signal was
measured using a Philips\footnote{Philips Photonique, Av. Roger
  Roacier, B.P. 520, F-19106 Brive, France} XP\,2020
photomultiplier and analyzed by an 
Analog-to-digital converter (ADC). The ADC was triggered
 by a threefold coincidence of signals coming from a 5\,mm thick
 plastic scintillator mounted behind the fibers using two Philips XP\,1911 photomultipliers for readout and from a second photomultiplier
XP\,2020 coupled to the second coupling piece of the fiber sample. The
light output measurement was performed before and after
irradiation. In addition the light output of the
non-irradiated scintillator reference samples and
light attenuation of the light guide
reference samples were regularily measured
%. Non-irradiated
%reference samples were used for comparison with irradiated samples
 to minimize systematic errors.\\

\section{Results}

From in-situ observations of proton and electron excited spectra
no remarkable difference could  be found \cite{lit6}. Consequently, we report here representative
results for both charge carrier excited spectra.\\
As described in \cite{lit5}, \cite{lit6}
all in-situ measured spectra show a two stage decay of the
scintillating light intensity in dependence on the energy dissipation
(or irradiation time).
%\cite{lit5}, \cite{lit6}.
In Fig.5  a typical degradation process for an electron excited
fiber  is presented with the wavelength as parameter. At the beginning
of the irradiation a similar time constant for all wavelengths can be
observed. A faster decay appears for
higher energy dose values of about 1\, Mrad in dependence on different
wavelengths of emission spectra. The short-wave emission shows the fastest
degradation in time (see also Fig. 2 in \cite{lit5}).

A recovery of the damaged fibers could be observed already during the
in-situ measurements. A considerable increase in light output was
observed  several times during the irradiation procedure after switching off the beam for only three minutes  (see Fig. 3 in \cite{lit5}). 

Exciting the same fibers by UV-light in the laboratory a few hours
after irradiation a long term recovery was measured. After 40 hours a
SCSF-78M fiber irradiated to 8.1\,Mrad showed 90 $\%$ of the light output
with respect to pre-irradiation (Fig.4 in \cite{lit5}). This process seems
however to depend on the fiber material and the integral dose (compare
Fig.2 in \cite{lit6}). 

The kind of excitation seems to be of particular importance for the
measured fiber light output. This will influence also the observed
recovery after irradiation and may explain the corresponding different
time constants for in-situ measurements and UV-excitation.

In a real experiment the scintillation light in fibers will be
produced by crossing charged particles. Therefore the multi-fiber test
samples were exposed to electrons from a Ru-source before and after
beam irradiation to measure light output and transmission. Indeed a
 different behaviour was found. As described in
\cite{lit1},\cite{lit3} the strongest damage was observed only about 
30 hours after
irradiation with a dose of 1\,Mrad for both light emission and
transparency with a complete recovery after two days (see Fig. 4 of
\cite{lit3}). 

A dose of 1\,Mrad, expected for the inner tracker of the HERA-B experiment
within one year, was placed to the above test samples within a few
minutes. This may have influenced the observed results in an inadmissible
way in particular if oxygen diffusion is important for damage and
recovery. To be closer to the  experimental conditions
% a study took place where 
the irradiation of scintillator and light guide
samples was performed up to a dose of 1.4 Mrad within 70 days. Half of the
fiber samples were covered by glue. The irradiation was done in air
and in nitrogen atmosphere. The results are displayed in Figs. 6 and
7.

Measurements going on for about half a year, using many times the same
samples, are difficult to perform keeping systematical errors small due
to some instability of the setup in time and mechanical damages of the
fragile samples. To minimize those effects,  non-irradiated
samples were measured every time in addition. All results are presented as ratios
of irradiated to non-irradiated fibers R$_S$ and R$_L$ for scintillators and
light guides, respectively. The maximum errors of these ratios have
been estimated to be about 30\,$\%$ including effects which may arise from
sample production.

How the irradiation was going on in time is demonstrated in Fig.
6a. The corresponding damage and recovery of four fiber samples is
shown in fig 6b. No effect could be observed outside the 30 $\%$ error
band. Neglecting the measurement errors and relying on the
pre-irradiation data points some damage 
%is observed 
may have happened
up to the maximum
dose followed by a long term recovery. The damage seems to be smaller
for glued fibers in particular in nitrogen atmosphere. Non-glued
fibers seem to recover only partly. 
%However all these effects are not really significant.

From Fig. 7a it can be seen that clear fibers were only irradiated up
to a dose of 400\,krad. For all measurements they were coupled to
scintillating fibers which were excited by electrons from a
Ru-source. Also here a maximum error of 30\,$\%$ has to be kept in mind
for the ratio R$_L$ of irradiated to totally non-irradiated (clear plus
scintillating) fibers shown in Fig 7b. Neglecting the error band no
damage is seen for the clear fiber irradiation itself. However
irradiating the scintillator to more than 1\,Mrad caused a decrease of
 the light
output from a coupled clear fiber to more than one half, i.e. more
than observed for the scintillating sample alone. After two weeks
complete recovery was found. The behaviour is the same for KURARAY and
pol.hi.tech. clear fibers in air and nitrogen.

\section{Summary}  

Several radiation hardness tests were performed for KURARAY
scintillating fibers SCSF-78M and clear fibers from KURARAY and
pol.hi.tech. Using high current proton and electron beams the
irradiation was performed both with very high and low dose rates. 

In-situ observations demonstrated a strong damage of scintillating
fibers for high dose rate exposures.  Both light emission and
transparency were decreased down to 20\,$\%$  for 1 Mrad. Short  and
long time recovery effects followed the irradiation.

For low dose rate conditions closer to a later experiment, a 30\,$\%$
decrease of scintillating fiber light output could not be excluded
recovering after three weeks. No significant influence of the fiber
coverage and the atmosphere during irradiation was found.

Clear fibers are apparently not damaged for doses up to 400\,krad. Coupled
to irradiated scintillating fibers the effective damage of the system
seems to increase.

\section*{Acknowledgement}

The fiber irradiation tests were possible only due to the kind support
of the Hahn-Meitner-Institute Berlin. In particular we want to thank 
the ISL accelerator team.

\newpage

{\bf Figure captions}\\

Fig. 1 :  Sketch of a multifiber test sample with indication of the
irradiation and measurement positions. The signal is measured via
coupling piece S1, the coupling piece S2 is used for the extraction of
a trigger signal. \\

Fig. 2 :  Fiber sample for in-situ measurements with irradiation
position and connections of fibers to  spectrometers. LGF: Light guide
fiber, SP1, SP2: Coupling to spectrometers 1, 2.\\

Fig. 3 :  Irradiation setup of the high dose rate proton and electron
irradiation. The arrows give the positions of plastic detectors for
beam profile measurement according to Fig. 4.\\

Fig. 4 :  Beam profile measured using polyvinylalcohol methylene blue
plastic detector. The figures belonging to each curve correspond to
the arrows in Fig. 3 giving the positions of such plastic films in the
irradiation setup: 1 - behind the exit window, 2- before the aperture
1, 3 - between aperture 1 and 2, 4- behind aperture 2, 5 - before the
Faraday cup, 6 - behind the Faraday cup.   \\

Fig. 5 :  In-situ measurement of the emission spectra of a
scintillating fiber excited by high dose rate electron irradiation in
dependence on the irradiation time in seconds.\\

Fig. 6 :  a.) Time dependence of irradiation dose for scintillating
fibers, \\
 b.) ratio R$_S$ of light output from irradiated to non-irradiated
 scintillating fiber samples in dependence on measurement time with
 respect to the first irradiation.\\

Fig. 7 :  a.) Time dependence of irradiation dose for scintillating
fibers and light guides, b.) ratio R$_L$ of light output from irradiated 
to non-irradiated clear and scintillating fiber samples in dependence
of the measurement time with respect to the first irradiation.
K$_L$: Light guide fiber from KURARAY, K$_S$: Scintillating fiber from
KURARAY, P$_L$: Light guide fiber from pol.hi.tech.\\
%
% \end{document}
%
%%%%%%%%%%%%%%%%%%%%%%%%%%%%%%%%%%%%%%%%
%      Figures
%%%%%%%%%%%%%%%%%%%%%%%%%%%%%%%%%%%%%%%%
% 
% \documentclass[a4,12pt,german] {article}
%\topmargin -2.0cm
% 
% \usepackage{a4,epsfig,rotating,amssymb}
% \headheight 0.0cm
% \pagestyle{empty}
% \textheight 23.0 cm
% \textwidth 15.5 cm
% \oddsidemargin 0.0cm
% \evensidemargin 0.0cm
% \begin{document}
%
\setlength{\unitlength}{1cm}
\begin{figure}
\vspace*{2.0cm}
\begin{center}
\epsfig{file=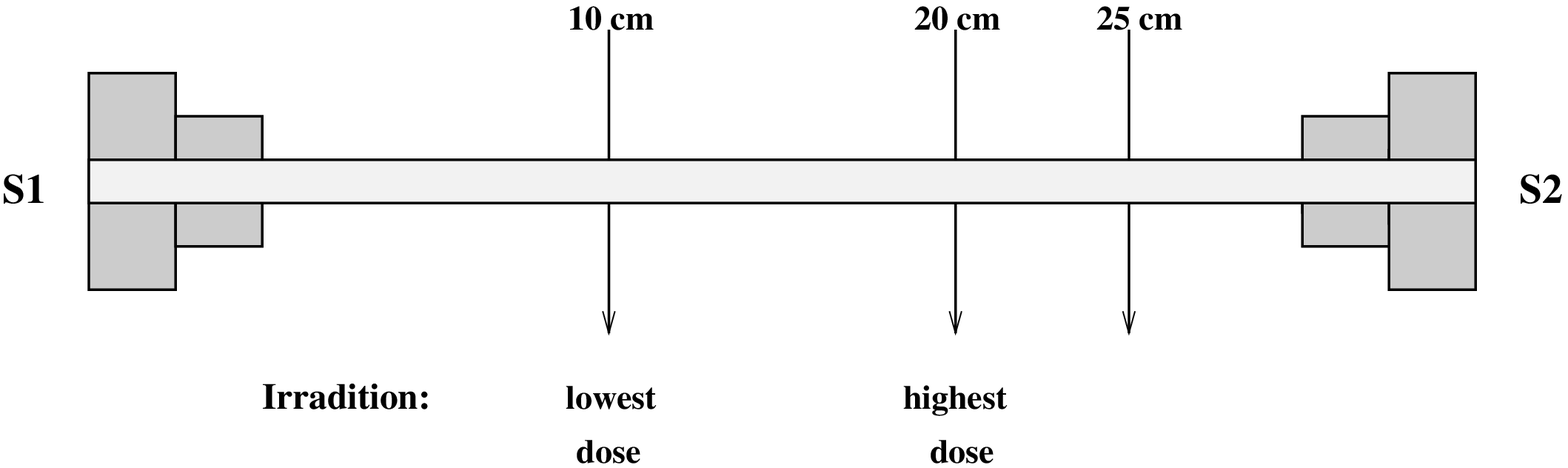,width=12cm}
\vspace*{0.5cm}
\end{center}
 \caption {Sketch of a multifiber test sample with indication of the
irradiation and measurement positions. The signal is measured via
coupling piece S1, the coupling piece S2 is used for the extraction of
a trigger signal.  }
\end{figure}

\setlength{\unitlength}{1cm}
\begin{figure}
\vspace*{2.5cm}
\begin{center}
\epsfig{file=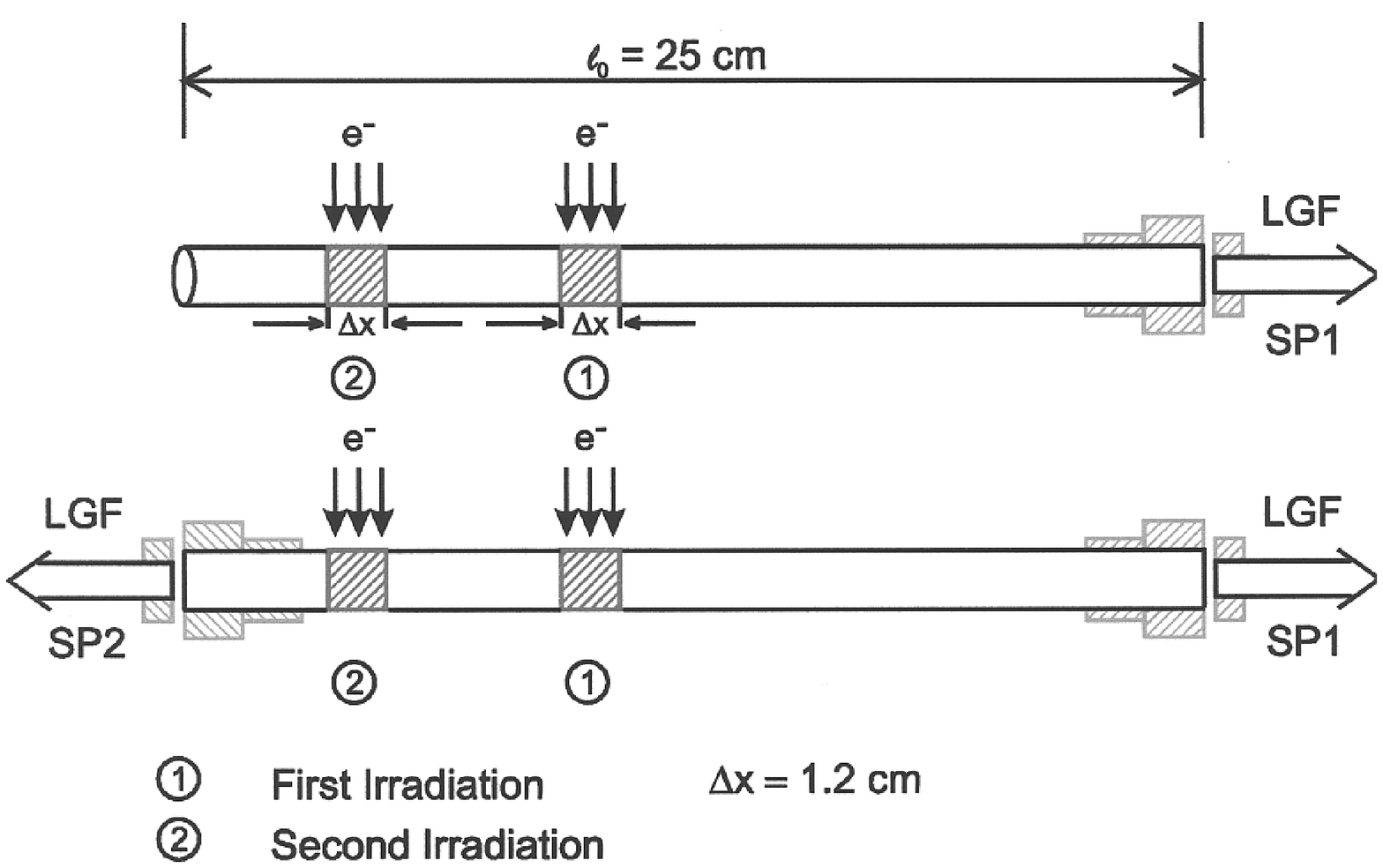,width=12cm}
\vspace*{0.5cm}
\end{center}
 \caption {Fiber sample for in-situ measurements with irradiation
position and connections of fibers to  spectrometers. LGF: Light guide
fiber, SP1, SP2: Coupling to spectrometers 1, 2.}
\end{figure}

\clearpage

\newpage

\setlength{\unitlength}{1cm}
\begin{figure}
\begin{center}
\epsfig{file=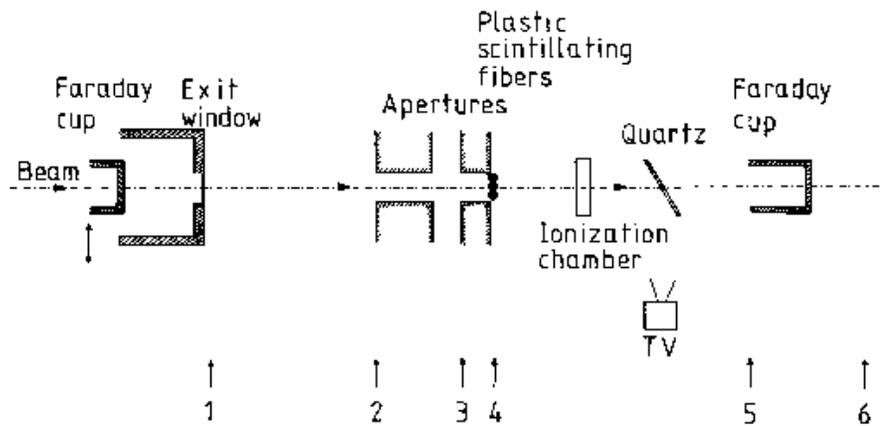,width=12cm}
\end{center}
  \caption{Irradiation setup of the high dose rate proton and electron
irradiation. The arrows give the positions of plastic detectors for
beam profile measurement according to Fig. 4. }
\end{figure}

\setlength{\unitlength}{1cm}
\begin{figure}
\vspace*{2.5cm}
\begin{center}
\epsfig{file=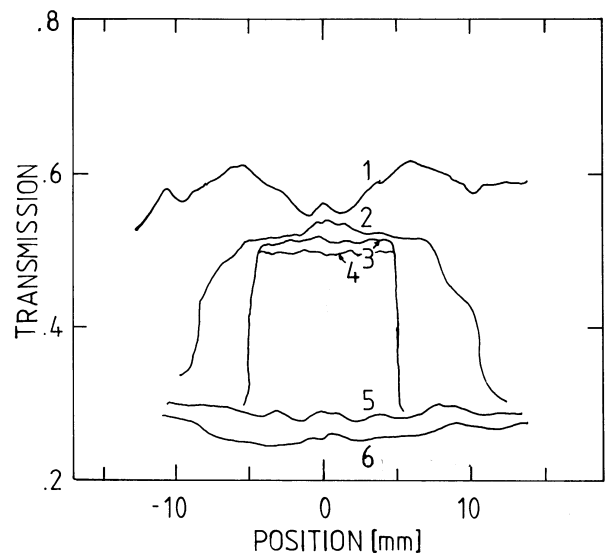 ,width=8cm}
\end{center}
  \caption{Beam profile measured using polyvinylalcohol methylene blue
plastic detector. The figures belonging to each curve correspond to
the arrows in Fig. 3 giving the positions of such plastic films in the
irradiation setup: 1 - behind the exit window, 2- before the aperture
1, 3 - between aperture 1 and 2, 4- behind aperture 2, 5 - before the
Faraday cup, 6 - behind the Faraday cup. }
\end{figure}

 \setlength{\unitlength}{1cm}
\begin{figure}
\begin{center}
\epsfig{file=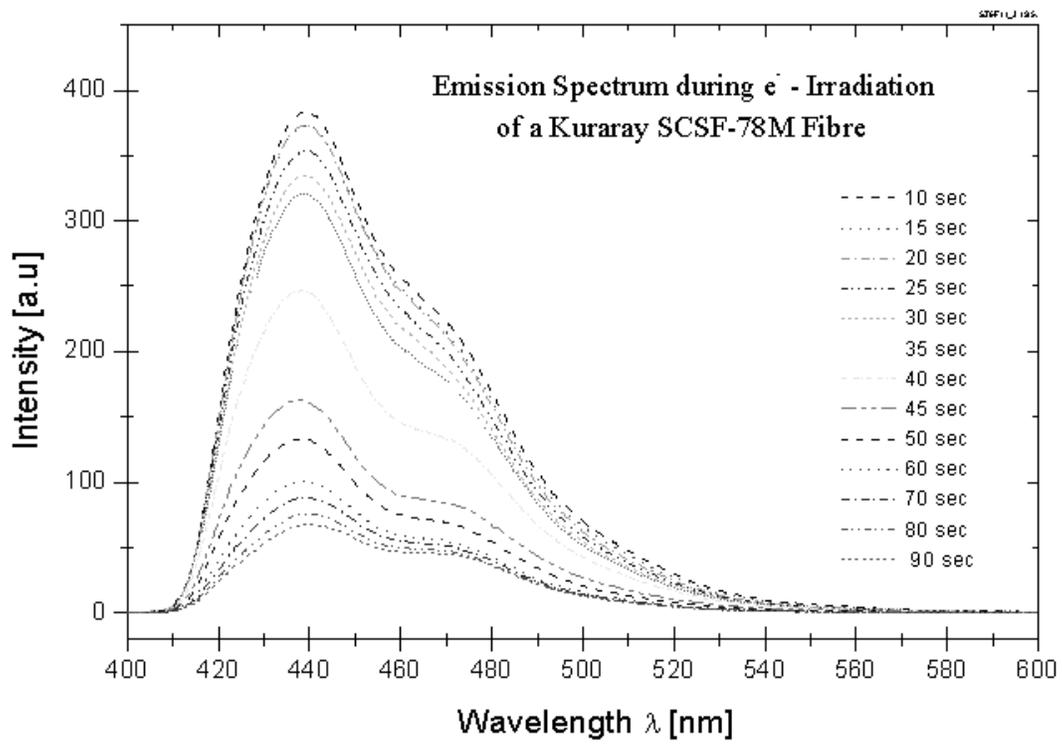,width=15cm}
\vspace*{1.0cm}
\end{center}
  \caption{In-situ measurement of the emission spectra of a
scintillating fiber excited by high dose rate electron irradiation in
dependence on the irradiation time in seconds. }
\end{figure}

\clearpage

\newpage

\begin{figure}
\vspace*{-2.0cm} 
%% \mbox{\epsfig{file=abb6.eps,width=15cm}} 
\mbox{\epsfig{file=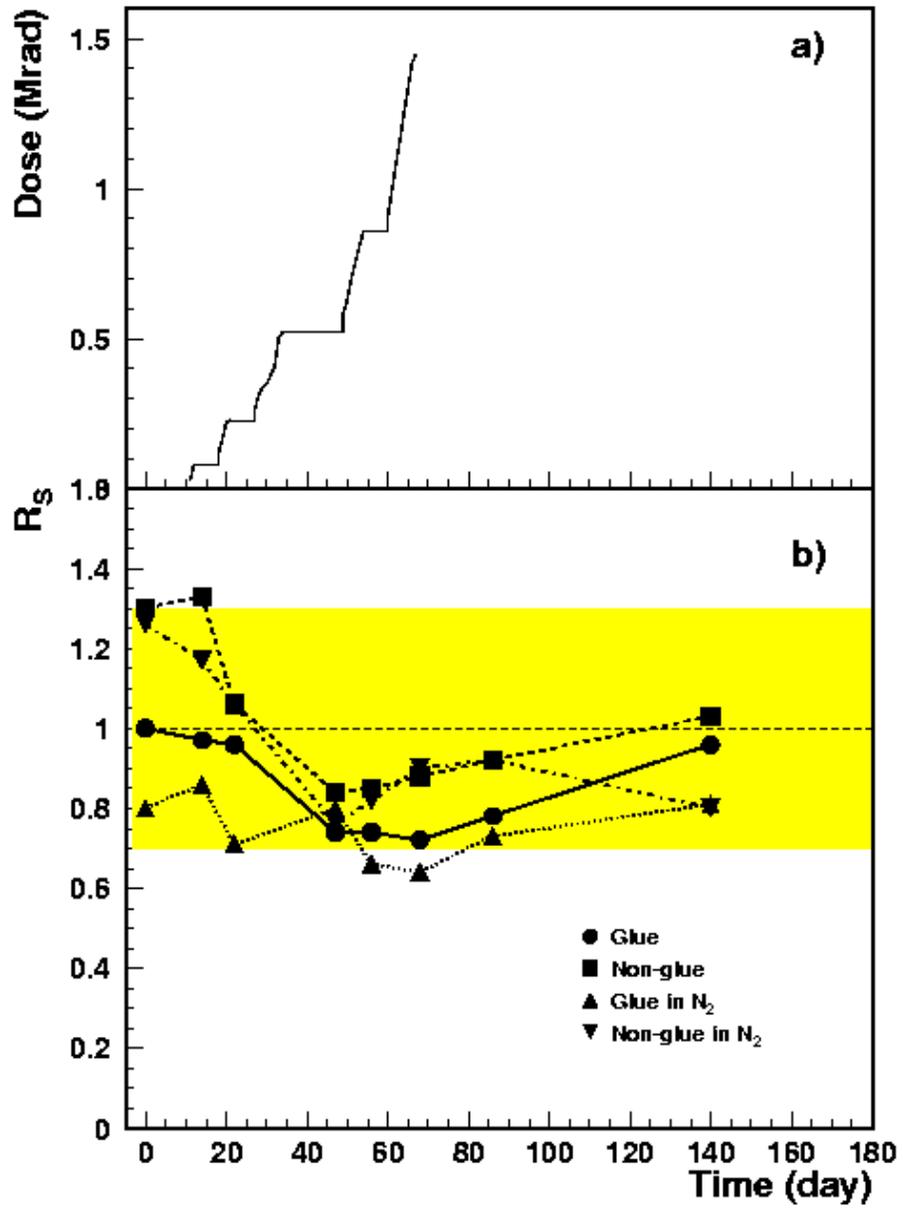,width=15cm}}
  \caption{ a.) Time dependence of irradiation dose for scintillating
fibers, 
 b.) ratio R${_S}$ of light output from irradiated to non-irradiated
 scintillating fiber samples in dependence on measurement time with
 respect to the first irradiation.}
\end{figure}
\clearpage

\newpage

\begin{figure}
\vspace*{-2.0cm}
% \mbox{\epsfig{file=abb7.eps,width=15cm}}
\mbox{\epsfig{file=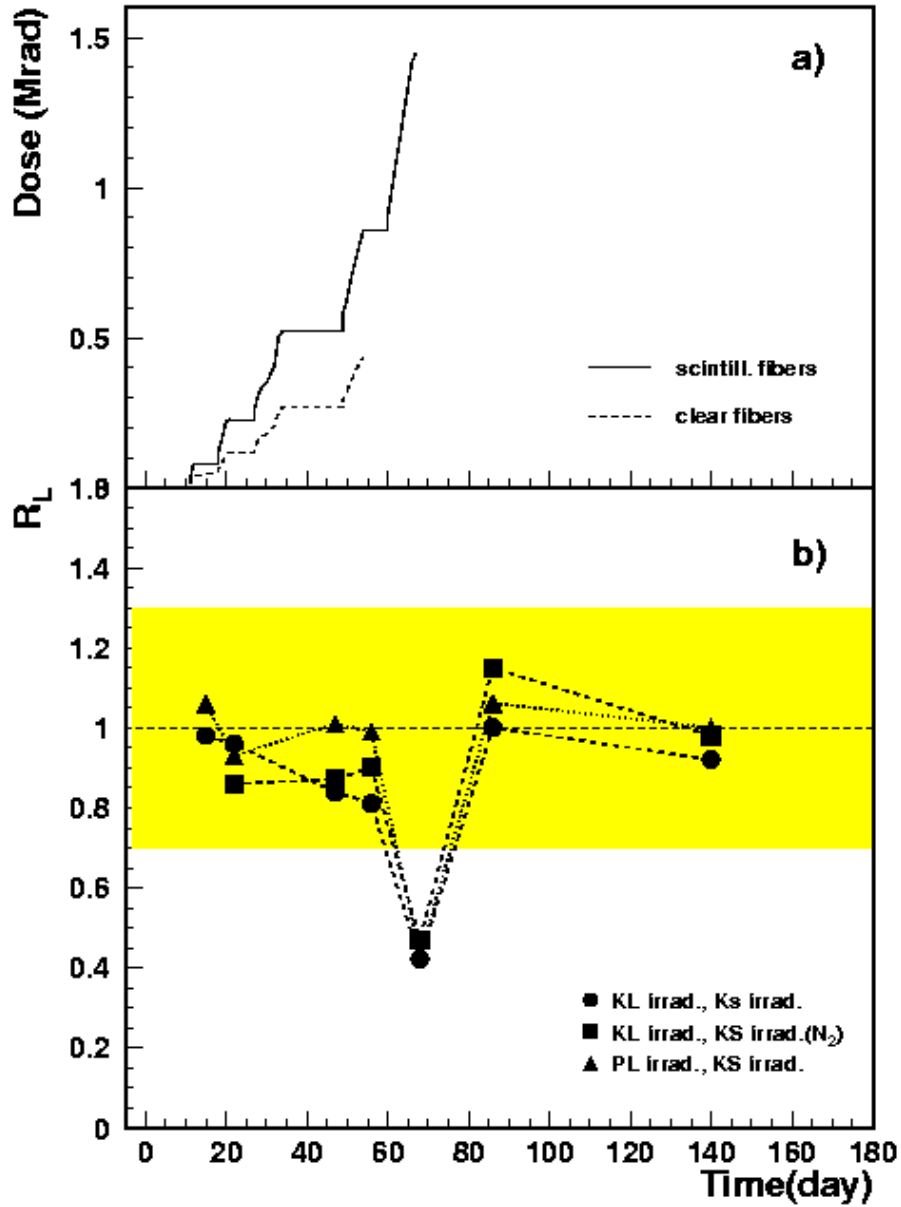,width=15cm}}
  \caption{a.) Time dependence of irradiation dose for scintillating
fibers and light guides, b.) ratio R$_L$ of light output from
irradiated
to non-irradiated clear and scintillating fiber samples in dependence
of the measurement time with respect to the first irradiation.
K$_L$: Light guide fiber from KURARAY, K$_S$: Scintillating fiber from
KURARAY, P$_L$: Light guide fiber from pol.hi.tech. }
\end{figure}

\end{document}